\documentclass{iau}

\usepackage{amsmath}
\usepackage{graphicx}
\usepackage{multirow}

\begin{document}

\lefttitle{Zs.~K\H{o}v\'ari et al.}
\righttitle{Magnetic activity under tidal influences in V815\,Her}

\jnlPage{1}{7}
\jnlDoiYr{2021}
\doival{10.1017/xxxxx}

\aopheadtitle{Proceedings IAU Symposium}
\editors{A. V. Getling \&  L. L. Kitchatinov, eds.}

\title{Magnetic activity under tidal influences in the 2+2~hierarchical quadruple system V815\,Herculis}

\author{Zs.~K\H{o}v\'ari$^{1,2}$,
        K.~G.~Strassmeier$^{3}$,
        L.~Kriskovics$^{1,2}$,
        K.~Ol\'ah$^{1,2}$,
        T.~Borkovits$^{1,4,5}$,      
        B.~Seli$^{1,2}$,
        K.~Vida$^{1,2}$,
        \'A.~Radv\'anyi$^{6}$}
\affiliation{{$^{1}$} Konkoly Observatory, HUN-REN Research Centre for Astronomy and Earth Sciences, Budapest, Hungary}
\affiliation{{$^{2}$} CSFK, MTA Centre of Excellence, Budapest, Hungary}
\affiliation{{$^{3}$} Leibniz-Institute for Astrophysics, Potsdam, Germany}
\affiliation{{$^{4}$} Baja Astronomical Observatory of University of Szeged, Baja, Hungary}
\affiliation{{$^{5}$} HUN-REN-SZTE Stellar Astrophysics Research Group, Baja, Hungary}
\affiliation{{$^{6}$} Moholy-Nagy University of Art and Design Budapest, Hungary}

\begin{abstract}
Tidal forces in close binaries and multiple systems that contain magnetically
active component are supposed to influence the operation of magnetic dynamo.
Through synchronization the tidal effect of a close companion helps maintain
fast rotation, thus supporting an efficient dynamo. At the same time, it can
also suppress the differential rotation of the convection zone, or even force the
formation of active longitudes at certain phases fixed to the orbit. V815\,Her
is a four-star system consisting of two close binaries orbiting each other, one
of which contains an active G-type main-sequence star. Therefore, the system
offers an excellent opportunity to investigate the influence of gravitational effects
on solar-type magnetic activity using different methods.
\end{abstract}


\maketitle

\section{V815\,Her: a single-lined multiple system with a chromospherically active primary}

\begin{figure}[tb]
  \centering
  \includegraphics[width=0.65\textwidth]{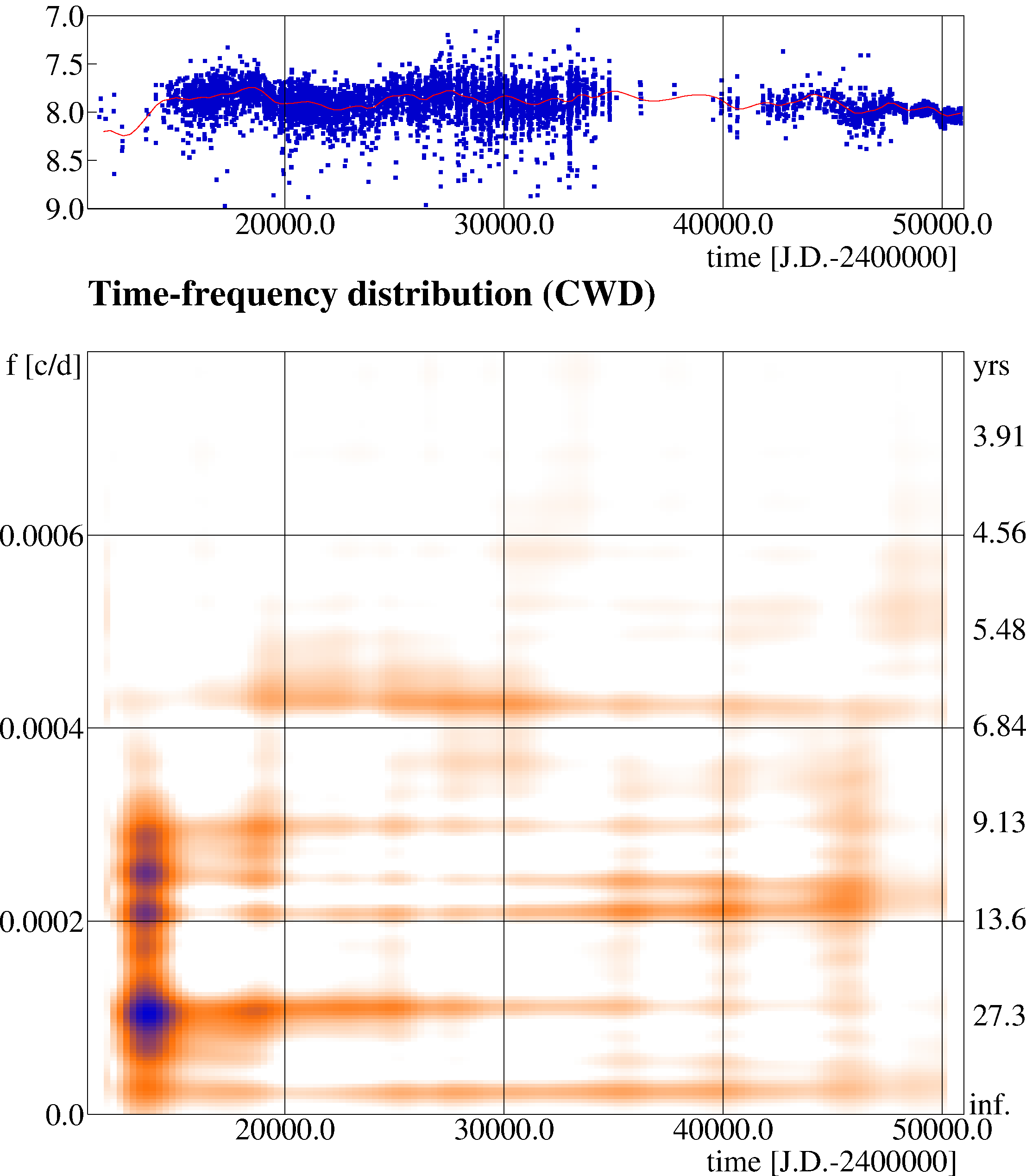}
  \caption{Long-term photometry and time-frequency analysis of V815\,Her.}
  \label{fig1}
\end{figure}

V815\,Her is a single-lined multiple star system with a G5-6 V type active primary component. It has long been known that the G star features cool spots on its surface. The photometric behavior of the system is dominated by the magnetic activity of the primary component, as shown by the long-term photometric data in Figure\,\ref{fig1}. For the period analysis we used our Time Frequency Analyzer package TiFrAn \citep{2009A&A...501..695K} with Choi-Williams distribution kernel. Several long-term brightness changes can be attributed to the chromospheric activity of the G star. The century-long archival data from Digital Access to a Sky Century @ Harvard (DASCH) database supplemented with photoelectric data from \citet{2000A&A...362..223J} reveals a slowly increasing cycle with steady amplitude around $\sim$6.5 years, two other cycles of about $\sim$9.1 and $\sim$13 years, while the longest feature points towards a timescale around $\sim$24-26 years with weakening amplitude (see the bottom panel of Fig.\,\ref{fig1}). The cycle lengths of $\sim$13 and $\sim$26 years are perhaps harmonics of $\sim$6.5 years.

The rapidly rotating ($P_{\rm rot}=1.8$\,d) G star (=V815\,Her~Aa) together with an unseen, but probably late M-dwarf component (=V815\,Her~Ab) form the close binary subsystem V815\,Her~A with $P_{\rm orb}=1.8$\,d, i.e. orbiting synchronously. The binary was also known to have a distant companion with an orbital period of about 5.7 years \citep{2005AJ....129.1001F}.

\section{V815\,Her~B: an eclipsing binary subsystem formed by two M type dwarfs}

\begin{figure}[h]
  \centering
  \includegraphics[width=0.5\textwidth]{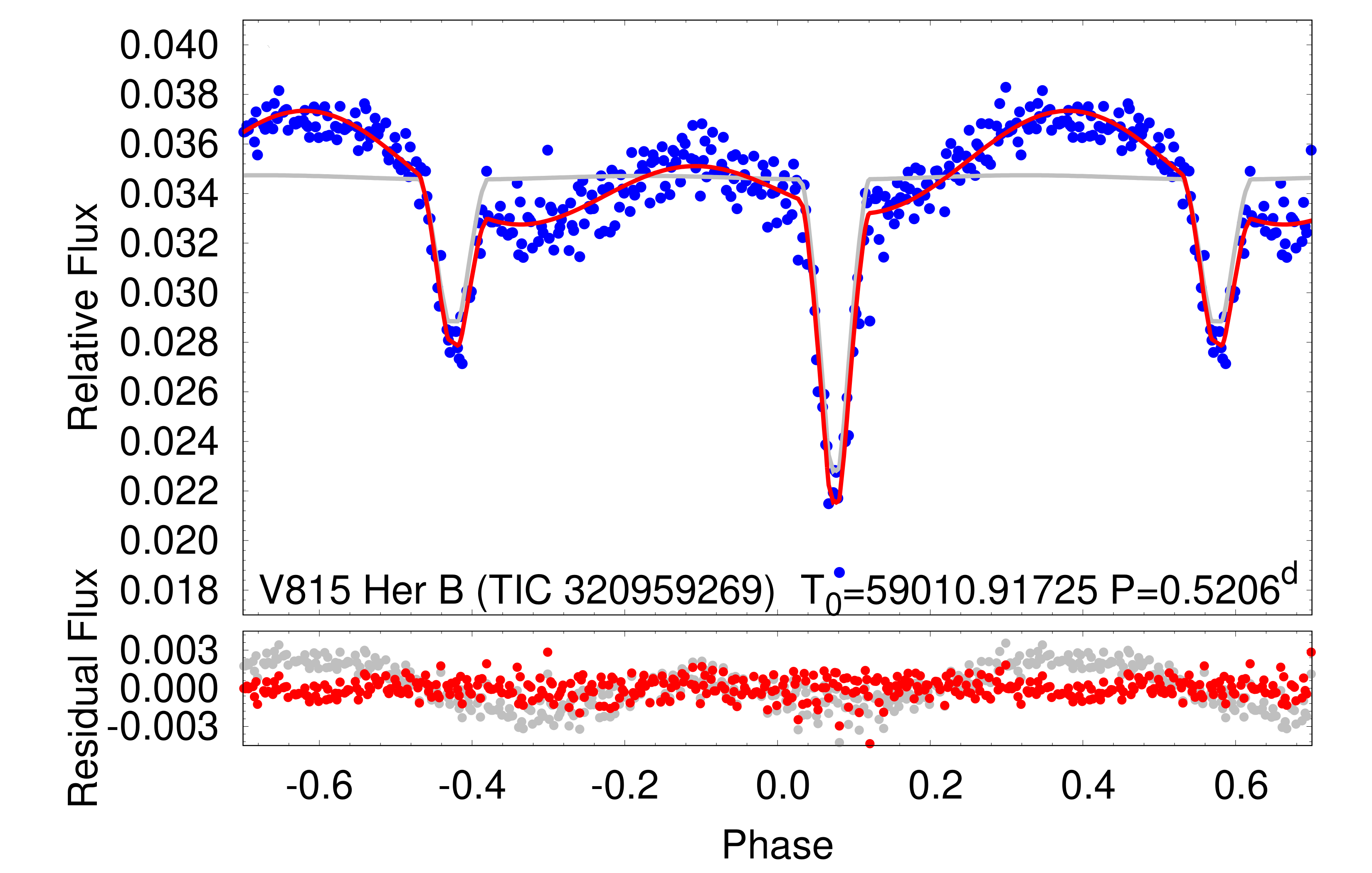}
  \caption{Cleaned and folded \emph{TESS} Sector 53 light curve
(blue) for V815\,Her~B with the orbital solution (grey) and the fitted spot model (red). Below are the residuals.}
  \label{fig2}
\end{figure}

Until our discovery, it was not known that the distant third body V815\,Her~B is actually itself an eclipsing close binary, whose members Ba+Bb are late M dwarfs (most probably M1+M5). In other words, the V815\,Her system is essentially a 2+2 hierarchical quadruple.
Figure\,\ref{fig2} shows the \emph{TESS} light curve of the primary and secondary minima of the V815\,Her~B eclipsing binary subsystem as an example. The light curve extracted from \emph{TESS} Sector~53 observations is folded with the 0.52-day orbital period of the subsystem. The orbital solution without assuming surface inhomogeneities are drawn with grey line, while the model drawn with red line fits also the light curve variations arising from stellar spots on the surface of one or both M dwarf components.

\begin{table}[tb]
 \centering
 \caption{Astrophysical parameters for V815\,Her~Aa}\label{tab1}
 {\tablefont\begin{tabular}{@{\extracolsep{\fill}}ll}
    \midrule
Spectral type           & G6 V \\
Gaia distance [pc]    &  $32.087\pm0.127$ \\
$V_{\rm br}$    [mag]          & 7.56 \\
$(B-V)$     [mag]         & 0.71 \\
$M_{\rm bol}$     [mag]        & $4.90\pm0.03$ \\
Luminosity [${L_{\odot}}$]         & $0.87\pm0.03$ \\
$T_{\rm eff}$ [K] &      5582$\pm$63 \\
$\log g$ (in cgs) &      4.27$\pm$0.04 \\
$v\sin i$ [kms$^{-1}$]    &       $30.0\pm1.5$  \\
$P_{\rm rot}=P_{\rm orb}$ [d]   &           $1.80983433$  \\
Inclination  [$^{\circ}$]            &       $75\pm5$   \\
Radius      [$R_{\odot}$]           &      $1.1\pm0.1$    \\
Mass          [$M_{\odot}$]           & $\lesssim$1.0   \\
Microturbulence  [kms$^{-1}$] & $1.14\pm 0.19$ \\
Macroturbulence  [kms$^{-1}$] & 4.2  \\
Metallicity [Fe/H] &  0.06$\pm$ 0.03  \\
Age          [Myr]           & $\sim$30   \\
    \midrule
\end{tabular}}
\end{table}

\section{Doppler imaging and differential rotation of V815\,Her~Aa}

\begin{figure}[thb]
  \centering
  \includegraphics[width=0.65\textwidth]{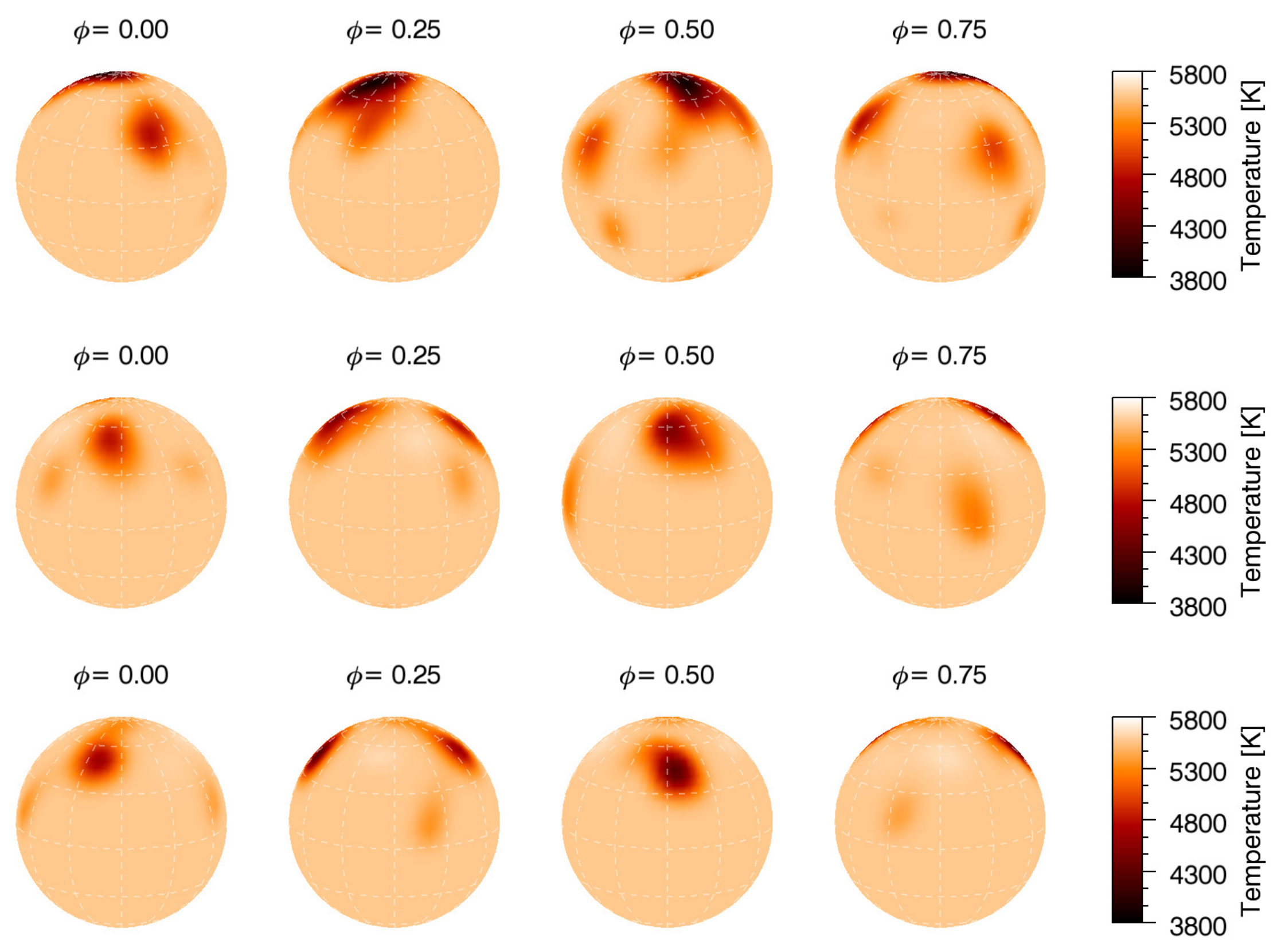}
  \caption{Three time-series Doppler images of V815\,Her~Aa from 2018.}
  \label{fig3}
\end{figure}

We performed a Doppler imaging study for the G star in order to reconstruct its spotted surface. For this we used 545 high-resolution optical spectra taken during a 9-month long observing run in 2018. The spectroscopic observations were carried out with the 1.2-m STELLA-II telescope of the STELLA robotic observatory \citep{2010AdAst2010E..19S} at the Iza\~na Observatory in Tenerife, Spain. From the spectra we created 19 independent Doppler images in time series (adopted stellar parameters are listed in  Table\,\ref{tab1}). For the image reconstruction we used the state-of-the-art Doppler imaging code \emph{iMap} \citep{2012A&A...548A..95C}. The images follow each other at intervals of 11 days on average. As an example, in Figure\,\ref{fig3} we show three surface maps, the 8th, the 9th and the 10th from the time series which indicate a constantly changing surface structure on a time scale of a few weeks. The rotational phases are calculated using the orbital period from \citet{2005AJ....129.1001F} according to the following equation:
  \begin{equation}
    HJD = 2450204.5802 + 1.80983433 \times E.
  \end{equation}

\begin{figure}[tb]
  \centering
  \includegraphics[width=0.48\textwidth]{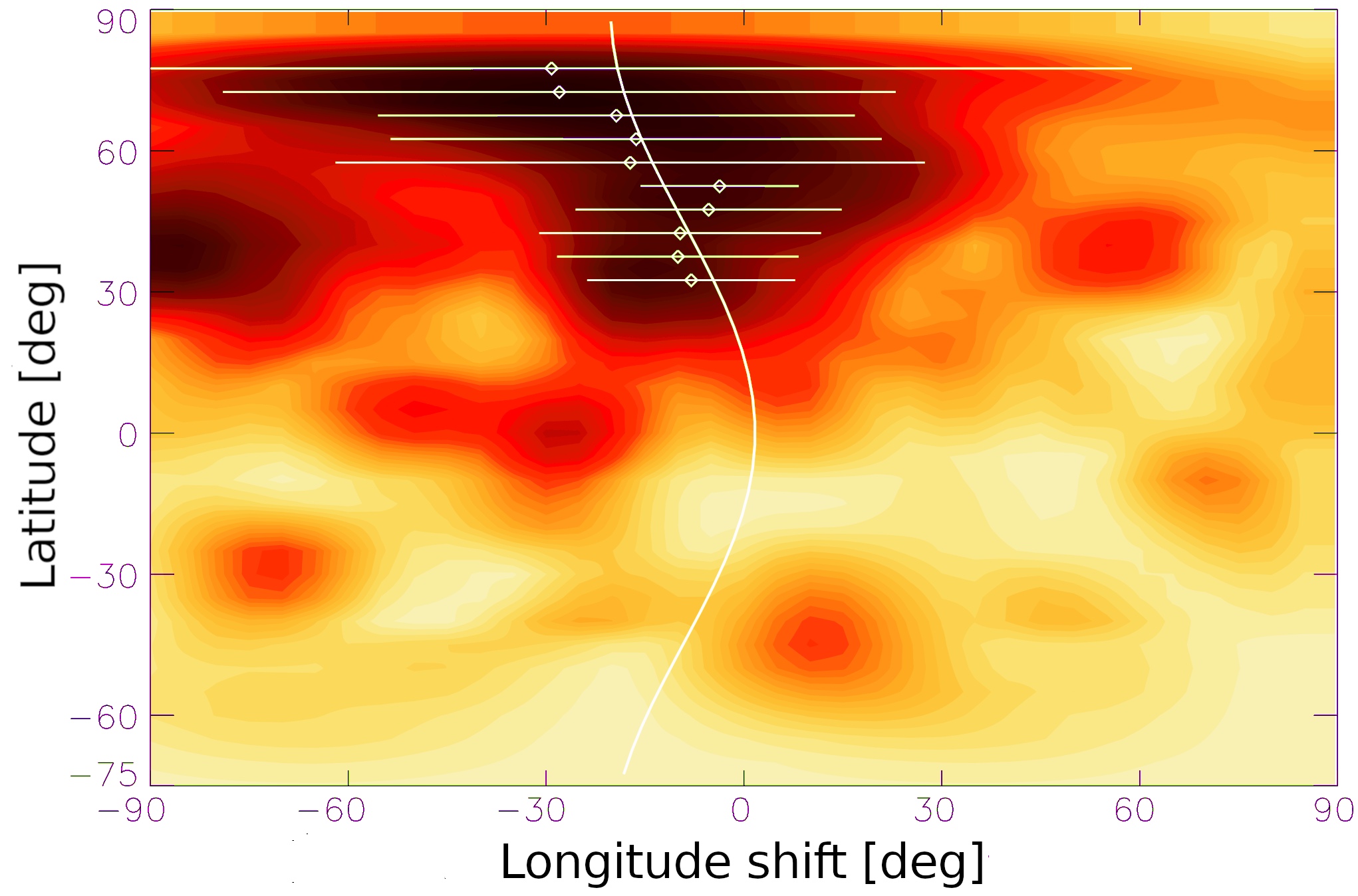}\includegraphics[width=0.46\textwidth]{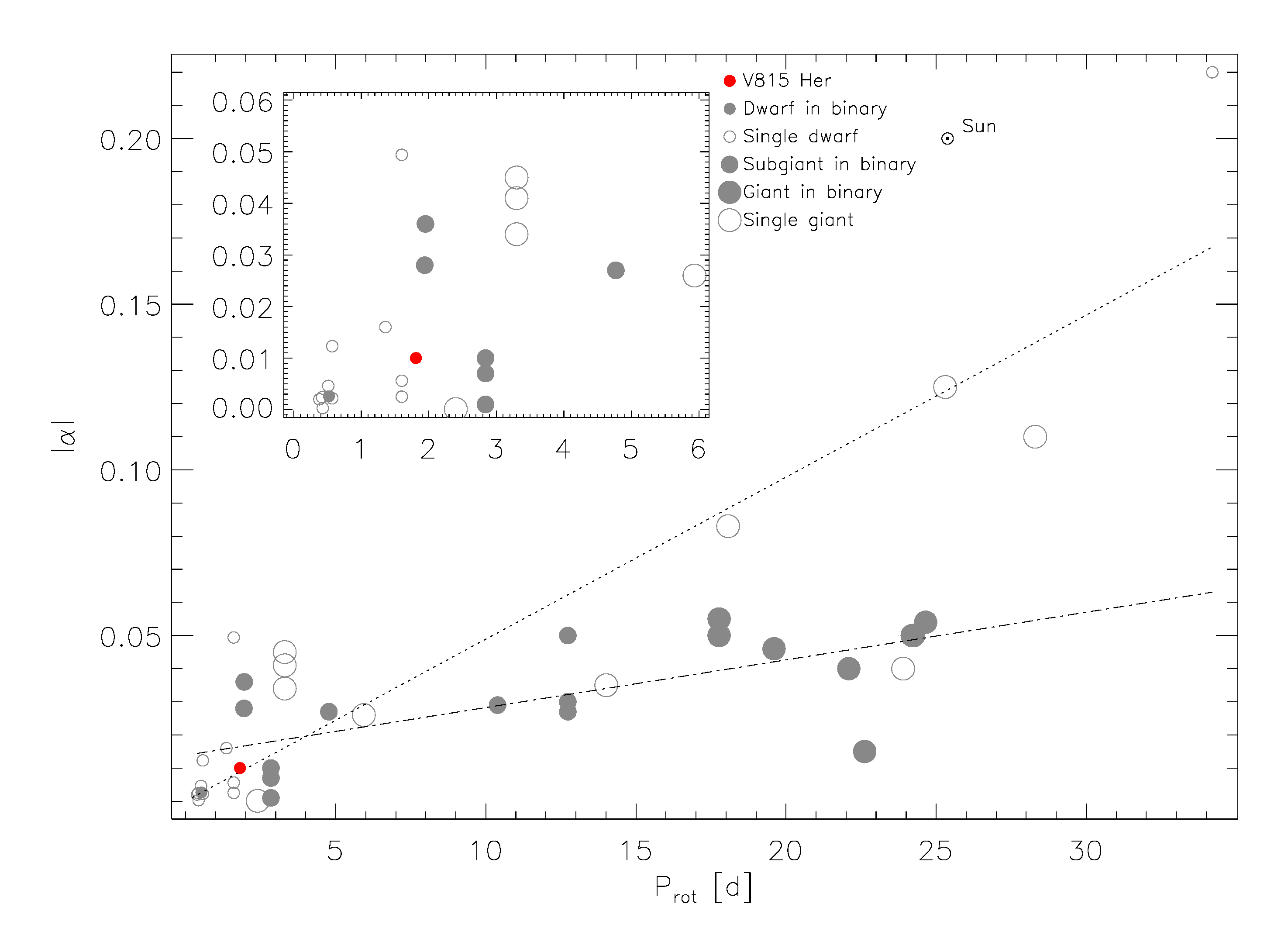}
  \caption{\emph{Left}: The average cross-correlation function map fitted with a solar-type rotation function. \emph{Right}: Surface shear parameter ($\alpha$) vs. rotation period ($P_{\rm rot}$) graph for active spotted stars.}
  \label{fig4}
\end{figure}

The latitude dependent surface rotation can usually be inferred from the cross-correlation of two successive Doppler images. Here we use ACCORD \citep{2012A&A...539A..50K,2015A&A...573A..98K}, a technique which enables the averaging of latitudinal cross-correlations in the case of a sufficient number of pairs of Doppler images not too distant in time, this way suppressing the effect of randomness and amplifying the correlation pattern attributed to surface differential rotation. In the left panel of Figure\,\ref{fig4} we plot the average cross-correlation function map from 18 individual cross-correlations (i.e. for each consecutive image pairs). The figure indicates how much longitude shift occurs at a given latitude due to surface shear during $\sim$11 days (which is the average time difference between the consecutive Doppler images). The fitted rotation function suggests a weak solar-type surface differential rotation with a dimensionless shear parameter $\alpha=(\Omega_{\rm eq}-\Omega_{\rm pole}) / \Omega_{\rm eq}=0.01.$

In the right panel of Figure\,\ref{fig4} we plotted an extended version of the surface shear parameter vs. rotation period graph ($|\alpha|$--$P_{\rm rot}$ graph) for active spotted stars based on \citet{2017AN....338..903K}. Open symbols correspond to (effectively) single active stars, grey symbols to active stars in binary (or multiple) systems. Symbol size increases from dwarfs to subgiants to giants. V815\,Her~Aa is represented by the red dot at $P_{\rm rot}$=1.8 days. The dotted and dash-dotted lines denote the linear fits for single stars and stars in binary systems, respectively, with slopes of $|\alpha|\propto 0.0049P_{\rm rot}$[d] and $|\alpha|\propto0.0014P_{\rm rot}$[d]. The difference clearly indicates that binarity does play a role in confining differential rotation of the convective envelope.

\section{Summary}

From the century-long archival photometric data (DASCH) of V815\,Her we found
a strong signal of a slowly increasing cycle length around 6.5 years on average, close to the 5.73-yr period of the wide (AB) orbit (see Figure\,\ref{fig1}). At this point we interpret this slowly changing, $\sim$6.5-year photometric period as the activity cycle of V815\,Her~Aa, which is perhaps triggered along the eccentric \citep[$e$=0.77, see][]{2005AJ....129.1001F} wide orbit.

We also demonstrated that V815\,Her, previously known as a close binary system with a `third body' in a wide orbit, and later suspected as a \emph{TESS} planet-host candidate, is actually a 2+2 hierarchical quadruple system of two close binaries, one of which (i.e., V815\,Her~B) is an eclipsing binary (see Figure\,\ref{fig2}).

We used high-resolution optical spectra from the STELLA robotic observatory taken during the 2018 observing run to create time-series temperature maps (Doppler images) of the spotted surface of the G component V815\,Her~Aa. Consecutive maps in Figure\,\ref{fig3} reflect significant spot activity and spot redistribution on a time scale comparable to the rotation period. From the latitudinal cross-correlation maps we measured a weak solar-type surface differential rotation
on the G star. The small shear parameter of $\alpha$=0.01, equivalent with $\Delta\Omega$$\approx$2$^{\circ}$/d surface shear, suggests that the differential rotation of the G star is probably suppressed (see Figure\,\ref{fig4}). This is consistent with our previous finding \citep{2017AN....338..903K} that tidal forces in close binary systems can indeed suppress the differential rotation of a component with convective envelope.\\

\noindent{\it Acknowledgements}
Authors from Konkoly Observatory acknowledge the Hungarian National Research, Development and Innovation Office (NKFIH) grants OTKA K-131508 and KKP-143986. LK acknowledges the NKFIH grant OTKA PD-134784. LK and KV are Bolyai János research fellows. KV is supported by the Bolyai+ grant \'UNKP-22-5-ELTE-1093, BS is supported by the \'UNKP-22-3 New National Excellence Program of the Ministry for Culture and Innovation from the source of the National Research, Development and Innovation Fund. STELLA robotic observatory was made possible by funding through the State of Brandenburg (MWFK) and the German Federal Ministry of Education and Research (BMBF). The facility is a collaboration of the AIP in Brandenburg with the IAC in Tenerife.

\bibliographystyle{iaulike}
\bibliography{kovarietal_iaus365proc}

\end{document}